\documentstyle[11pt,aaspp4]{article}

 \newcommand{\kms}{km s$^{-1}$}

\lefthead{Miriam Centurion et al.}  \righthead{Nitrogen abundances in
Damped Ly$\alpha$ absorbers}

\begin{document}

\title{Nitrogen abundances in Damped Ly$\alpha$ galaxies}

\author{Miriam Centuri\'on}
\affil{Instituto de Astrof\'\i sica de Canarias, 38200, La Laguna, 
Tenerife, Spain} 

\author{Piercarlo Bonifacio, Paolo Molaro, Giovanni Vladilo} 
\affil{Osservatorio Astronomico di
 Trieste, Via G.B.  Tiepolo 11, I-34131, Trieste, Italy}

%\affil
%\altaffilmark{1}

\begin{abstract}

Nitrogen lines  of the NI 1134\AA\ and 1200\AA\ multiplets 
in the Damped Ly$\alpha$ (DLA) galaxies 
at $z_{\rm abs}$ = 2.309, 2.827  and 3.025 
toward the QSOs 0100+1300, 1425+6039 and  0347$-$3819 
respectively, have been detected by means of high resolution spectra 
(R $\sim$ 2 $\times$ 10$^4$) obtained
with  4-m class telescopes at ESO (La Silla, Chile) and ORM (La Palma, Spain).
The two NI multiplets offer a considerable range in oscillator strengths
and the possibility of disentangling
Ly$\alpha$ interlopers.
The derived nitrogen abundances
for the three damped systems  are 
[N/H]= --2.68$\pm$0.11, --1.57$\pm$0.09, --2.07$\pm$0.13, respectively.
The behaviour of
nitrogen  relative to iron-peak and $\alpha$-elements has been
investigated by considering all the
extant NI determinations for a total of 9  DLA galaxies. 
We have estimated  the fraction
of iron locked into dust grains   to convert the
observed [N/Fe] ratios into  
overall (dust plus gas) relative abundances, [N/Fe]$_{corr}$.  
The ratios [N/$\alpha$] have been mostly 
determined by using sulphur as a tracer of $\alpha$-elements
which is unaffected by dust. 
The [N/Fe] and [N/$\alpha$] ratios show high dispersions,
of one order of magnitude or more, which have no equivalent in
other element-to-element ratios  in DLAs.   
The lowest values of the [N/Fe]$_{corr}$ and  [N/$\alpha$] 
ratios are at variance with the values  measured in Galactic halo 
stars of
similar metallicity, suggesting that part of the DLA galaxies
do not follow the chemical evolution of the Milky Way. 
The DLA nitrogen abundances and their dispersion show  some
similarities
with those observed in dwarf galaxies.
Comparison with chemical evolution models show that
the lowest [N/Fe]$_{corr}$ and [N/$\alpha$] DLA values are close to what would be 
expected for a pure secondary origin of nitrogen, 
while higher values are mostly consistent with a primary component. 
The behaviour of nitrogen abundance ratios  
can be ascribed, in general, to the delayed release of nitrogen 
in the course of evolution.
However it is difficult to conciliate this interpretation 
with the lowest [N/$\alpha$] values measured, since an expected 
enhancement of $\alpha$-elements respect to the iron-peak elements is not
observed simultaneously in these DLA galaxies. 
In two cases, relatively high [N/$\alpha$] values are observed which 
require also a more complex chemical evolution to be explained.

\end{abstract}


\section{Introduction}

Damped Ly$\alpha$ (DLA) systems are complexes of neutral gas  
detectable in the spectra of background
quasars as HI absorption lines with large column densities, N(HI)$>$
10$^{20.3}$ atoms cm$^{-2}$ (Wolfe et al. 1986).
The damped Ly$\alpha$ absorptions are found at all redshifts    
$0.1 < z_{\rm abs} < 4.5$ and are associated with metal lines of
low ionization and often  high ionized species as well.

Although the exact nature of the DLA systems has yet to be clarified, there
is consensus on the fact that they are  the progenitors
of present-day galaxies.  Damped galaxies might be massive rotating
disks which will evolve to the present-day spiral galaxies (Wolfe et
al.  1986, 1995a, Prochaska \& Wolfe 1997) or less massive objects 
progenitors of dwarf galaxies
(Tyson 1988, Pettini et al.  1990, Meyer and York 1992, Steidel et al.  1995).
Understanding which interpretation is the more appropriate has important
implications for any theory of galaxy formation and evolution. 
Imaging in the field of the background quasars
suggests that more than a single population of galaxies give rise to
the DLAs, even though the results are sometimes ambiguous, especially
at high redshift (Briggs et al.  1989, Steidel \& Hamilton 1992).
At $z \leq$ 1 the candidate associations include gas-rich galaxies
with a variety of morfological types (Le Brun et
al.  1997; Lanzetta et al.  1997). Very recently Rao \& Turnshek (1998)
discovered  two DLA
absorbers at very low redshift (z=0.091, z=0.221) and by means of
imaging observations they exclude that they are luminous
spiral galaxies.
 
DLA galaxies have low metallicities, typically between 
1\% and 10\% of solar,  a signature of the
enrichment of the first stellar generations.
The precision obtained in abundance studies of DLAs is remarkable and  
comparable to that attained in the interstellar studies of our own
Galaxy, thus offering important clues for understanding the
nature of these objects. 
Element abundance ratios are  tracers of galactic
chemical evolution  
(see e.g. Lauroesch et al. 1996, Matteucci, Molaro \& Vladilo, 1997), 
however the results obtained for DLA
galaxies 
are subject to some debate.
The typical overabundance of $\alpha$-elements compared to iron-peak elements
($\alpha$/Fe) found in Galactic metal-poor stars is observed in 
DLAs by means of the Si/Fe ratios, which is interpreted as a signature of 
the first stages of
chemical evolution of a spiral galaxy like the Milky Way (Wolfe et al. 1995b,
Lu et al. 1996).  
On the other hand, the $\alpha$/Fe ratios in DLAs are observed close to solar 
when a dust-free
diagnostic such as S/Zn 
is used (Molaro, Centuri\'on \& Vladilo 1998; hereafter MCV), or when 
the observed 
abundances of refractory elements (as Fe and Si) are 
corrected by dust depletion (Vladilo 1998).

Nitrogen is an important element when attempting to understand the chemical history of
DLA galaxies. The nucleosynthesis of N, 
which takes place mainly in low and intermediate mass 
stars, is quite complex and can be used to probe conditions typical of early
stages of galactic chemical evolution. 
Although the dominant
processes which lead to the enrichment of N are still not fully
understood it seems clear  
that secondary synthesis of N -- which occur in stars of all masses --  plays an
important role at high metallicities (Vila-Costas
\& Edmunds 1993). In this case secondary origin of N refers to its
production in the CNO cycle from seed C and O nuclei created in
earlier generation of stars.  However, at low metallicity levels,
nitrogen observations in halo stars (Carbon et al.  1987, Tomkin \&
Lambert 1984) and in  HII regions of dwarf irregular
galaxies (Thuan, Izotov \& Lipovetsky 1995; Kobulnicky \& Skillman
1996) require a significant primary component in the synthesis of this
element.  Primary production of N is obtained in intermediate mass stars 
when freshly synthesized
C in the helium-burning shell penetrates into the hydrogen-burning
shell of the same star where it is converted into N during the Asymptotic 
Giant Branch phase of their
evolution (Renzini \& Voli, 1981).  Even though a general 
consensus is lacking,
it has been suggested that
primary N might be also produced --
esentially by the same mechanism -- in massive stars of low metallicities 
(Woosley \& Weaver, 1995). 
This production is strongly dependent
on the assumed treatment of convection in stellar interiors but it seems   
necessary in order to explain the high N abundances relative to O
observed in some
dwarf irregular galaxies and some DLAs (Matteucci, Molaro \& Vladilo, 1997
and references therein).

A further advantage of N is that it is not depleted onto dust in
the Galactic interstellar medium (Savage \& Sembach 1996; Mayer,
Cardelli \& Sofia 1997); therefore  N abundances measured in 
DLA systems and in extragalactic
HII regions are expected to reflect the real (gas plus dust) N abundances.

The first attempt to detect nitrogen in DLAs was made by
Pettini, Lipman \& Hunstead (1995) who derived a stringent  upper limit
([N/H]$<$ --3.15) for the absorber at $z$=2.279 towards QSO
2348--1444.  The upper limit on the [N/O] ratio in this system ($<$ --1.24)
implied the lowest ratio ever found either in stars or galaxies at low
metallicity, in line with a secondary origin of nitrogen.  Pettini
and colleagues  argued that this result is consistent with a delayed delivery
of N created mainly in intermediate mass stars compared to O
enrichment produced in short-lived massive stars
(Edmunds \& Pagel 1978; Garnett 1990).

Nitrogen was first detected in the damped absorbers at  
z=1.776 towards MC 1331+170 (Green et al.  1995; Kulkarni et al.
1996) and at z=3.390 towards
QSO 0000-2620 (Vladilo et al.  1995; Molaro et al.  1996).  In both 
cases the reported N/O abundance ratio  is much higher than
the
upper limit derived by  Pettini, Lipman \& Hunstead (1995). These results 
gave the first indication of an
intrinsic dispersion of nitrogen abundances in DLAs, as pointed 
out by Vladilo et al.  (1996) and Molaro et
al.  (1996). The high nitrogen
abundance ratios are not easy to understand and non conventional chemical
evolution models of the type used for dwarf irregular galaxies have
been proposed by Matteucci, Molaro \& Vladilo (1997).

In a subsequent work, Lu et al.  (1996) found a
stringent upper limit in the DLA at $z$=2.994 towards QSO 1946+7658,
in line with the result reported by Pettini and colleagues.  Lu et al.
(1996) 
%argue that the low N/O upper limits is a consequence of %Type
%II supernovae nucleosynthesis (responsible for the bulk %of O
%production therefore low N/O ratios) and they 
interpreted this low N/O value 
as an evidence of the  first
stages of chemical evolution of a spiral galaxy like the Milky Way.
More recently, four new nitrogen abundances were obtained by 
Lu, Sargent \& Barlow 
(1998) offering further evidence for an intrinsic dispersion of the
nitrogen abundances in DLAs.  From the analysis of the [N/Si] ratios in 
DLA galaxies the latter argue that the observed dispersion can be expected 
considering
the time delay between oxygen and nitrogen delivery.

In this paper we describe three new NI abundance determinations   in
the DLA systems
at $z$ = 2.309 towards QSO 0100+1300,  
at $z$ = 2.827 towards QSO 1425+6039,
and  at $z$ = 3.025 towards QSO 0347$-$3819. 
The detections are based on 
the six NI lines of the multiplets at 1134\AA\, and 1200\AA.
A preliminary report for the DLA system towards QSO 0347$-$3819 was
given in Vladilo et al.  (1996).  
In Section 2 and 3 we present our observations and   column
density measurements which also include SII, SiII, OI, and FeII determinations.   
Abundance measurements for all the DLAs 
with nitrogen detections
are compiled from our data and from the literature,
and are presented in Section 4. 
Element-to-element ratios in particular N/Zn, N/Fe, and N/S are  discussed 
in Section 5 
in connection
to the origin of N in the framework of chemical evolution models.  
The implications for our understanding of the nature of DLA
galaxies are  summarised in Section 6.

\section{Observations and data reduction}

As part of a search for nitrogen absorptions in DLAs we
obtained optical spectra of QSO 0100+1300 (V=16.6), QSO 1425+6039
(V=15.8), and QSO 0347--3819 (V=17.3). 
The spectra of the first two QSOs
were obtained with the Utrecht Echelle Spectrograph (Walker \& Diego 1985)
at the Nasmyth focus of the 4.2m William Herschel Telescope of the
Observatorio del Roque de los Muchachos on La Palma island.  The
spectra of QSO 0347--3819 were obtained with the CASPEC
spectrograph at the cassegrain focus of the ESO 3.6m telescope at La
Silla, Chile (Pasquini \& D'Odorico 1989).  In both spectrographs an
echelle grating of 31.6 grooves/mm and a 1024x1024 Tektronik CCD with
square pixels of 24 $\mu$m in size were used.  The CCD was binned at a
step of 2 pixels along the dispersion and the slit width was set at 2.1"
(CASPEC) and 2.2" (UES) in order to project the slit onto 2 binned pixels of
the detector.  The slit width
matched the seeing  
which was around 2 arcsecs during the observations. 
The full width at half maximum of the instrumental profile, $\Delta
\lambda_{instr}$, was measured from the emission lines of the
Thorium-Argon lamp spectra recorded frequently during the
observations.  The resulting resolving power R= $\lambda$/$\Delta
\lambda_{instr}$ was R $\simeq$ 27000 (UES spectra) 
and 19500  (CASPEC spectra), corresponding to a velocity
resolution of $\Delta$v $\simeq$ 11.5 \kms\ and 15.4 \kms,
respectively.
The journal
of the observations is given in Table 1.

The spectra taken on different nights were reduced separately and then
averaged using weights according to the square of their S/N ratios.
Cosmic ray removal, sky subtraction, optimal order extraction and
wavelength calibration were performed using the ECHELLE routines
implemented within the MIDAS software package developed at ESO.   
Typical internal errors in the wavelength calibrations are of
$\simeq$ 0.01-0.02 \AA. 
The wavelength scale of the spectra was
corrected to vacuum
heliocentric scale.  Finally, each order of the spectra was
normalised using a spline to connect smoothly the regions free from
absorption features.
The signal-to-noise ratios of the final  spectra --
estimated from the $rms$ scatter of the continuum near the absorptions
under study -- are typically in the range between 10 and 20. 
For the two DLAs at higher redshift, the OI 1302\AA, OI 1355\AA\ and 
SiII 1304\AA\
transitions fall in a  part of the spectrum where the 
signal-to-noise ratios can be as high as 45 thanks to the
increase of the CCD quantum efficiency towards the red.

\section{Column densities}

Column densities have been derived 
by best fitting theoretical Voigt profiles
to the observed absorption lines  
via $\chi^2$ minimization.
This step was performed   using the 
routines FITLYMAN (Fontana \& Ballester 1995) included in the MIDAS
package.
Before the fit, the theoretical profiles were convolved   
with the instrumental point spread function
determined from the analysis of the emission lines
of the arcs. 
Portions of the profiles contaminated
by intervening Ly $\alpha$ absorbers
were excluded from the fit. 
The FITLYMAN routines determine column densities,
broadening parameters ($b$ values) and redshifts of the
absorption components, as well as the fit errors for
each   of these quantities. The laboratory wavelengths of the 
transitions investigated are listed 
in Table 2 together with the
oscillator strenghts adopted for computation of
theoretical profiles.

When several unsaturated transitions are detected 
for a given element, the column density is
estimated by applying the fit procedure both to the individual
transitions and to the full set of available lines. 
No differences were found between the mean column density 
resulting from
individual fits and the column density obtained from the 
simultaneous fit
of all the available lines. 
In these cases we adopted the dispersion from individual measurements as
conservative column density errors.

Nitrogen column densities were determined from the
six transitions of the two NI multiplets at
$\lambda$1134\AA\ and $\lambda$1200\AA\ (see Table 2). 
The NI lines occur in the Ly $\alpha$ forest, but 
the identification of the absorptions is  
quite reliable because it is very unlikely that
each one of the six  transitions is   blended with a Ly $\alpha$
interloper.  The oscillator strengths span about one order of
magnitude offering a large dynamical range for the measurement  
of the column density.  
The detection of the faintest transitions of the $\lambda$ 1134\AA\
multiplet is especially important to  get rid of saturation effects.
These transitions are not saturated and therefore uncertainties in $b$(NI) 
do not affect the derived NI column densities.

The fits of nitrogen lines for the three DLA systems
under investigation are shown in Fig.  1, 2 and 3 and  the resulting
column densities are reported in Table 3.  In the same table
we list the rest of extant determinations of NI column densities or
upper limits taken from the literature.  In total there are nine NI detections
available: four from our project, one from Kulkarni et al.  (1996),
four from the recent measurements reported by Lu, Sargent \& Barlow   (1998).  In
addition there are eight upper limits (Pettini, Lipman \& Hunstead
1995; Lu, Sargent \& Barlow   1998).
 
In order to understand the nucleosynthetic origin of nitrogen it is
necessary to compare its abundance with that of other elements.
In addition to the nitrogen lines, we  therefore
searched  for transitions of important species such as OI, SiII, SII, and FeII
which fall in the wavelength range of our spectra (see Table 2).  
Column densities and $b$ values of these elements  
for the DLAs towards QSO 0347--3819 and QSO 1425$+$6039 are reported in 
Table 4 and   
for the system 
towards QSO 0100+1300 are
given in  MCV.

For saturated transitions, lower limits to the column density were estimated
from the equivalent width obtained from the best fit to the absorption.
For undetected transitions, upper limits to the column density  
were derived from  3$\sigma$ upper limits of the equivalent width.
In both cases the conversion from equivalent widths to column
densities was performed in the  optically thin limit (linear part of the
curve of growth).  

Oxygen column densities in DLAs are generally uncertain
by  more than one order of magnitude owing to the saturation of 
the OI 1302\AA\ line which is present even at low metallicities. 
In order to constrain the OI
column density we  fitted the OI 1302\AA\ line by adopting the
$b$ value obtained from the analysis of the NI multiplets, which is 
expected to trace the same neutral gas as OI. 
The $b$ value of both atoms should  in fact be very similar:
$b_{OI}$ = 0.94 $b_{NI}$ for pure thermal broadening and
$b_{OI}$ = $b_{NI}$  for pure turbulence. 
The other accessible OI transition at 1356\AA\ is extremely weak
(f$_\lambda$ = 1.248 x 10$^{-6}$) and is not detected in the DLA systems under study 
(see e.g. Fig. 4 and Fig.  5). The derived upper limits are given
in Table 4.

Measurements of SII column densities   are derived
from the SII triplet at 1254\AA\ which is generally unsaturated. 
This triplet, however, can be  
contaminated by the Lyman $\alpha$ forest and we have been able
to measure the sulphur column density only for two of the three 
DLA systems under study. 
As far as iron is concerned, there are six FeII transitions available,
listed in Table 2. For each of the three DLA galaxies
studied here, there are one or two FeII  
unsaturated lines, and free from blends, which have allowed us to derive 
reliable 
iron column densities.
Silicon transitions in the available spectral ranges 
are usually heavily saturated hampering attempts to derive accurate SiII 
column densities.  More details on 
the measurements performed in individual systems are given in the 
next sub-sections.

\subsection{System at $z_{\rm abs}$=2.309 towards QSO 0100+1300}

The NI column density of this system was measured via the
simultaneous fit of the two lines at shorter wavelength of the
NI 1134\AA\ multiplet. 
As one can see in Fig. 1, the reddest transition of this multiplet
is in fact contaminated by a  Ly $\alpha$ interloper.
The NI 1200\AA\ multiplet presents strong contamination and is
highly saturated. 

The column density that we derive -- log N(NI) = 14.69($\pm$0.07) --   
is at variance with the one found by Lu, Sargent \& Barlow (1998),
log N(NI)= 15.29 $\pm$ 0.35 (see Table 3).  
These authors  obtained this result by means of Keck HIRES data, 
but details on the measurement are deferred to a subsequent paper.
The large error bar reported by Lu, Sargent \& Barlow (1998) 
suggests that N(NI) was obtained from the 1200\AA\ multiplet.
In fact, the 1134\AA\ multiplet falls below the spectral range usually covered  
by HIRES observations (see Table 1 in Lu et al. 1996).    
We stress that our measure relies instead on the analysis of the
unsaturated NI absorptions at
$\lambda\lambda$ 1134.1 and 1134.4 \AA.

The main absorption component of the system towards QSO 0100+1300
falls  at $z_{\rm abs}$ = 2.30907 in our spectra.
A secondary component at $v$ $\sim$ 35 \kms\, has been also detected  
in the most intense absorption lines of this system (Wolfe et al. 1994, MCV). 
Analysis of the most intense NI lines 
does not yield information on this component,
owing to Ly$\alpha$ contamination.
However, close inspection
of our spectrum redwards of the 1134.41\AA\ line 
shows a portion free from
Ly$\alpha$ contamination without traces of absorption
exceeding the noise level. Therefore the $v$ $\sim$ 35 \kms\, component --
if present in NI at all -- must be negligible compared to the
main one. 

Wolfe et al. (1994) in their higher resolution ($\Delta v$=8 \kms) spectra at
redder wavelengths (4427\AA\, - 6900\AA) revealed an asymmetric profile 
of the main
component suggesting the presence of an additional component at $v$
$\sim$ 8 \kms.  The effect of this double structure of the main
component on the column density of NI is negligible.  In fact, using a
two-cloud model the NI fit gives a total column density  
within 0.01 dex of the one-cloud model.
Abundances of other elements like SII, FeII, MgII, MnII, PII (from our data) and 
ZnII (from Wolfe et al. 1994) in this system 
are discussed in MCV.

\subsection{System at $z_{\rm abs}$=3.025 towards QSO Q0347--3819 }

The intense  SiII 1304\AA\ and OI 1302\AA\ absorptions
outside the Lyman $\alpha$ forest show two
components; the stronger lying at $z_{\rm abs}$= 3.02476 and the
weaker at $z_{\rm abs}$= 3.02562 (see Fig. 4).  The column 
density  of the secondary
component is less than 10\% of that of the main 
one for both species.  For NI,
FeII and SII, only the main component is clearly detected.  In these cases
we used a single component fit and adopted the resulting column
density as the total column density.

The NI multiplets are clearly identified, but some contamination from
Lyman $\alpha$ absorption is visible, particularly in the vicinity of
some lines of the 1134\AA\ multiplet.  To minimize  contamination we
measured the column density by fitting the central parts of the line
profiles.  The weakest line of the 1134\AA\ multiplet -- noisy and
apparently contaminated -- was excluded from the fit.  Results are  
shown in Fig. 2  and in Table 3.

The column density of SiII was estimated from the analysis of 
the 1304\AA\ line, which falls outside the Lyman
$\alpha$ forest and has the lowest oscillator strength among the
available SiII transitions. This fit
gives  log N(SiII) = 15.07 $\pm$ 0.43 
and  $b$ = 21 $\pm$ 5 \kms. 
An attempt to fit also simultaneously the 1193\AA\
transition, which appears to be unblended, 
yields instead log N(SiII) = 15.50 $\pm$ 0.69 and $b$ = 17 $\pm$ 4 \kms.
Silicon abundance must
obviously be treated with some caution. 

The analysis of NI and SiII yields consistent values of the broadening
parameter, although the value $b$= 17.9 $\pm$ 1.1 \kms\ derived from the NI
multiplets is more precise.  We adopted this value, with a
3$\sigma$ error ($\pm$ 3.3 \kms), to fit species for which the $b$
value is undetermined.

There is only one transition of the SII triplet which is not contaminated by Ly
$\alpha$ interlopers, namely the one at 1260\AA.  The identification
is supported by the perfect coincidence of  the redshift and by the
narrowness of the  absorption profile.  We obtain log N(SII) = 14.77 $\pm$ 0.06 where
the error is the fit error at a fixed  value of $b$.  An additional $\pm$
0.06 error results when the above mentioned range of possible $b$
values is taken into account. The SII column density is
fairly independent of the adopted value of the
broadening parameter.    

Many FeII transitions of this system are potentially present in
the spectral range of our spectra (see Fig. 4). However, all these lines
lie in the Lyman $\alpha$ forest and many of them are expected 
to be close to the noise level .  We clearly detect the 1144\AA\ line and a
hint of absorption in correspondence with other FeII transitions.  Our
analysis is mostly based on the 1144\AA\ line, but the synthetic
spectra were superposed to all the other lines for a consistency check.  
By excluding a few pixels of the violet wing of the
1144\AA\ absorption -- which we suspect is affected by Ly $\alpha$
contamination -- we obtain log N(FeII) = 14.35 $\pm$ 0.1 at $b$= 17.9
\kms.  The 1096\AA\ line gives a firm upper limit at log N(FeII) $\leq$
14.45, clearly indicating  that the stronger 1144\AA\ line cannot be
significantly saturated. By fitting the full absorption profile of the
1144\AA\ line without adopting  the above quoted
value of $b$, we obtain log N(FeII) = 14.45 at $b$ = 30 \kms.

We estimated the OI column density from the 1302\AA\ line.  By
constraining the range of broadening values between $b$= 14.6 \kms\
and $b$= 21.2 \kms\ we derive log $N$(OI) = 16.97 $\pm$ 0.39 and log
$N$(OI) = 15.61 $\pm$ 0.19, respectively.  The central value for the oxygen
column density given in Table 4  corresponds to the adoption of $b$= 17.9
\kms, our best estimate of the NI broadening parameter.

\subsection{System at $z_{\rm abs}$=2.82680 in QSO 1425+6039}

The NI column densities of this system were measured via the
simultaneous fit of  two lines   of the
NI 1134\AA\ multiplet. 
The reddest transition of this multiplet
is in fact contaminated by a  Ly $\alpha$ interloper (see Fig. 3).
The NI 1200\AA\ multiplet is clearly saturated and presents contamination 
in its weakest line.
From the analysis of the 1200\AA\ multiplet alone
Lu et al. (1996) derived  a lower
limit (log N(NI) $\geq$ 14.62) which is consistent with 
the result (log N(NI) = 14.70$\pm$0.04 provided here (cfr. Table 3).

The NI absorptions in our spectra occur at $z_{abs}$ =
2.82680.  Lu et al.  (1996) found an absorber sub-complex at $z_{abs}$
= 2.83058 ($\Delta v \simeq$ 300 \kms) with an HI column density at 5\% that 
of 
the main component. 
The shift in radial velocity of this secondary component
prevents contamination of the 1134 multiplet from
which we derived the NI column density.

In Fig. 5 we show the spectral regions of the SiII, SII, FeII and OI 
transitions.
Three out of the four SiII transitions available are severely blended with other
absorptions. Only the saturated SiII 1304\AA\, line could be fitted,  
providing a lower limit to the Si abundance as explained in Section 3.    

All three SII lines present contamination by the
Ly$\alpha$ forest making it impossible to estimate, or even to derive a
reliable upper limit to the sulphur column density in this DLA.

Only the 1143\AA\, and 1144\AA\, transitions  were
used to derive the FeII column density
since the other lines are
severely blended (see Fig.  5).  
The $b$ value derived from the iron lines (25.2 $\pm$ 1.8 \kms)
is in good agreement with that
obtained from the nitrogen lines (23.0 $\pm$ 1.8 \kms). 
The iron column density that we infer -- 
log N(FeII)= 14.45 $\pm$ 0.06 -- 
matches very well the result
reported by Lu et al.  (1996) for this system, namely log N(FeII)=
14.48 $\pm$ 0.04.

The OI 1302\AA\ transition is heavily saturated and blended. In this case 
we simply estimate an upper limit to the OI column density from the undetected 
1356\AA\ transition.

\section{Elemental abundances}

Table 5  lists measurements of the N, S, Si, O, Fe, and Zn abundances 
taken from our search and from the literature for all the DLAs with available nitrogen
abundance determinations. 
The elemental abundances are scaled to the solar
reference value by using the standard definition
[X/H] = log (X/H)$_{obs}$ -- log (X/H)$_{\odot}$.
The column density ratio (X/H)$_{obs}$ is obtained
by using the HI column densities given in Table 3. 
The solar system values adopted for all the elements discussed here 
are the meteoritic abundances 
reported by Anders \& Grevesse (1989), with the
exception of N and O for which we adopt the most recent solar phostospheric values   
given by Grevesse \& Noels (1993) (see Table 5).
The Zn  abundances are included in Table 5 because Zn  is a good indicator 
of the system's metallicity since it is practically unaffected by the presence of dust
(Pettini et al. 1994, 1997). 
The conversion from column densities to abundances requires several
logical steps which are briefly discussed in the following sub-sections.

\subsection{Ionization corrections}
 
DLA systems have large column densities -- N(HI) $>$ 2 x 10$^{20}$ cm$^{-2}$ --
and are optically thick to interstellar or intergalactic
ionizing photons  with h$\nu$ $>$ 13.6 eV in the absorber
rest frame. 
Detailed ionization calculations indicate that species with
ionization potential I.P. $>$ 13.6 eV,
such as NI, OI, SII, SiII, FeII, and ZnII, are 
the dominant ionization states and can be used to infer
directly the elemental abundance 
(Fan \& Tytler 1994; Lu et al.  1995).   
We assume that ionization corrections are negligible
for these species. 

\subsection{Contributions from intervening HII regions}
 
If the galaxy associated to the DLA system hosts
an intervening HII region, the ionized gas may produce
an extra contribution to the column densities of 
species with I.P. $>$ 13.6 eV.
This effect should not alter the abundances of species with IP very 
close to 13.6 eV such  
as OI and NI (since these species will be ionized in an intervening HII region)
but could increase the apparent abundances of other species with
higher ionization potentials, such as
SiII, SII, FeII, and ZnII. 
The net result would be an artificial decrease of the
[N/Fe] and [N/S] ratios and an increase of the apparent
metallicities [Fe/H] and [S/H] 
which are discussed in Section 5. 
The effect should  only be important when the intervening HII gas
has the same radial velocity and  
comparable column density of the HI region. 
Ionized regions with N(HII) $<$ 2 x 10$^{19}$ cm$^{-2}$
would not affect the abundance analysis in any case.
The intervening HII regions will have in general a different 
kinematics from the
HI gas and this will lead to differences in the radial velocity profiles of 
neutrals (originated only in HI gas) and singly ionized species
(originated both in HI and HII gas).
Therefore careful analysis of the absorption
profiles of metal lines should help  
disentangle the contribution of the ionized gas, if at all  present. 
From the analysis of our data we do not find systematic differences
between the $b$ values derived from neutral nitrogen and those derived from
singly ionized species.
Inspection of the highest quality spectra 
available (see e.g Lu et al. 1996) does not reveal systematic
differences between the profiles of neutral and low ionization species
suggesting that contributions from HII
regions are generally negligible.

\subsection{Dust depletion}

A fraction of the elements in DLAs are expected to be in the form of 
dust grains thus reducing the gas phase 
abundance which is precisely what we are measuring.
The evidence for dust in DLAs is manyfold.  One  
is the differential reddening of quasars with and without foreground
damped absorption (Pei, Fold \& Betchold 1991).  Another piece of evidence  comes
from the apparent enhancement of the Zn to Cr abundance ratio in DLA
systems (Pettini et al.  1994, 1997).  Based on the expectation that
Zn tracks Cr,  Pettini and colleagues  interpret the observed [Zn/Cr]
enhancement as a signature of  cromium depletion onto dust grains.
Vladilo (1998) calculated  dust-to-gas ratios in  
DLAs with Fe, Cr, and Zn measurements
and found quantitative agreement with the dust-to-gas ratios 
based on the reddening determinations by Pei, Fold \& Betchold (1991). 
The detection of the 2175\AA\ emission bump in  high redshift MgII
absorbers (Malhotra 1997) 
and the realization that type II SNae can produce dust 
(Danziger et al. 1991) also suggest  that dust is
present in the early stages of galactic evolution.   
Depletions in DLA absorbers are expected to be lower than in
the Galactic ISM since the dust-to-gas ratios in DLAs are
between $\simeq$ 2\% and $\simeq$ 25\% (Vladilo 1998) that of  the Galactic ones. 
However, care must be taken 
when converting column densities into abundances  
while considering  elements which are known to be
depleted from gas to dust.  

There are different pieces of evidence which  indicate that
nitrogen is not depleted in the interstellar medium.
From the theoretical point of view,   
the large activation energy of N$_{2}$ prevents  
involvement of nitrogen in the gas-phase reactions which lead to
dust formation in stellar atmospheres (Gail \& Sedlmayr 1986), and
hence interstellar accretion of N onto dust grains  
is not expected.  From the observational point of view, two recent works
confirm that interstellar N is not depleted into dust grains.
On the  one side, 
interstellar HST observations of the  NI $\lambda$1160\AA\ doublet  
(Meyer, Cardelli \& Sofia 1997) 
show that the gas-phase abundance of nitrogen does not
decrease in sight lines with higher H$_{2}$ fraction which are
indicative of self-shielding environments more hospitable to grains.  
On the other, ISO observations of the N-H stretch
spectral region at 2.96 $\mu$m 
limit the solid-state of N abundance to log (N/H)$_{dust}$ $< -6 $,
corresponding to a N depletion lower than 0.01 dex, for the   
line of sight to IV Cyg No.  12 which is known to be heavily reddened
(Whittet et al.  1997).   
 
Among the elements considered here, also O, S and Zn 
are almost unaffected by depletion
in Galactic interstellar clouds, their typical deficiency
with respect to solar values being in the range
[$-0.19,0.0$], [0.00], and [$-0.25,-0.13$] dex, respectively
(Meyer et al. 1998, Savage \& Sembach 1996, Roth \& Blades 1995). 
By contrast, refractory
elements such as Si and Fe are known to be  
depleted by 1 or 2 dex in the Galactic ISM.  In particular,
a fraction of iron as high as about  94$\%$ is locked in dust grains in
the warm gas of the Galactic disk, and a fraction even greater in cold
clouds (Savage \& Sembach 1996). These elements are expected to be 
depleted also in DLAs
and in Section 5 we discuss   how to estimate the amount of depletion.

\subsection{Reference abundances}

The measurement of dust depletion in the Galactic ISM relies on the
adoption of a suitable reference level for the elemental abundances.
The solar abundance pattern is usually adopted as a standard, but
there are indications that the   pattern observed in nearby B-type stars  
might be more appropriate (see Savage \&
Sembach 1996). Metal abundances are typically $-0.2$ dex lower
in B-type stars than in the solar system and some authors prefer to use
2/3 of the  solar
system values as a standard for "cosmic" abundances (see Savage
\& Sembach 1996, Mathis 1996).
Adoption of this abundance reference
would yield dust depletions consistently lower in absolute values.

The average interstellar abundance of nitrogen,   
$<$log(N/H)$>$$_{ISM}$ = --4.12$\pm$0.02 (Meyer, Cardelli \& Sofia 1997),
is slightly deficient with respect to the solar value 
log (N/H)$_{\sun}$=--4.03 $\pm$0.07 (Grevesse \& Noels 1993),
yielding an underabundance $<$[N/H]$>$$_{ISM}$ = --0.09$\pm$0.07.
Also nearby B stars show a similar underabundance, with
$<$log(N/H)$>$$_{Bstars}$ = --4.17$^{+0.11}_{-0.17}$ (Venn 1995). 
These results indicate non
depletion of N into interstellar grains if the B-type star reference
is adopted, or depletion of $\simeq -0.1$ dex if the solar pattern is
adopted.
Meyer et al.  (1997) pointed out that limitations in deriving
[N/H] in the Galactic ISM no longer lie  in the measurement itself,
but in the accuracy of the oscillator strenghts 
and in the solar photospheric determinations.  Due to these uncertainties  
the authors claimed that  
it is still premature to rule out an
interstellar N abundance either equal to, or 2/3 of solar.
We adopt here the solar reference pattern, and for nitrogen
the solar photospheric value 
of Grevesse \& Noels (1993) quoted above.  
This is not in sharp contrast with the fact that we assume  
nitrogen to be undepleted in DLAs because the expected absolute value for
Galactic depletion $<0.1$ dex  would become negligible 
at the low dust-to-gas ratios typical of DLA absorbers. 
Consequences of adopting the B stars as standard for the abundances
are considered in the next section.

\section{Discussion}

The absolute nitrogen abundances in DLA absorbers shown in Table 5
span more than two orders of magnitude,
with  --3.7 dex $<$ [N/H]  $<-1.6$ dex. 
From the point of view of chemical evolution models, relative
abundances are more reliable than absolute abundances because they generally
depend only on the elemental nucleosynthesis and stellar lifetimes,
whereas absolute abundances are affected by specific assumptions
underlying the models.   
Here we discuss the abundance
of nitrogen relative to the iron-peak and to $\alpha$ elements,
which are the main products of Type Ia SNae and Type II SNae,
respectively.   We compare measurements of [N/Fe] and [N/$\alpha$]
ratios
in DLAs and
in other astrophysical sites of similarly low metallicity,
namely Galactic halo stars, blue compact galaxies, and dwarf irregular
galaxies.
These abundance ratios are then compared with predictions of
galactic chemical evolution models.

\subsection{The [N/Fe] abundance ratio}

The nitrogen abundances  relative to iron measured in DLA systems
are plotted
versus [Fe/H] in Fig. 6.
The filled symbols represent our data
while open symbols denote measurements taken from the literature. 
The [N/Fe] values are
deficient compared to the solar ratio
and do not exhibit a   trend with metallicity.
The measurements   show a
relatively high amount of dispersion, with one value as low as [N/Fe]
= --1.3 ($z$=2.84 towards QSO 1946+7658) and several high values
around [N/Fe] = --0.3. There is also a hint of a possible higher dispersion 
at lower metallicities.

The dotted lines in Fig. 6 indicate
the range of  measurements in Galactic halo dwarfs with $ -3
\leq$ [Fe/H] $\leq 0$ (Carbon et al. 1987).  
The solid line
is the regression found by these authors  with a scatter
of $\sigma=\pm$0.19 dex. The large scatter  
in the stellar measurements  reflects  the
difficulties in deriving N abundances from NH bands.  
The non dependence of  [N/Fe] with stellar 
metallicity has been considered as evidence for a primary 
component of nitrogen
(Pagel 1985, Carbon et al. 1987).

The dot symbols  in Fig.  6 represent  the  measurements in Blue
Compact Galaxies (BCG) taken from the sample of Thuan,
Izotov \& Lipovetsky (1995). BCGs abundances  take up
about the same region of metal-poor stars. However,
most of them lie
below the linear regression of stellar data and closer to 
DLA galaxies.  At the lowest metallicities there are
no BCGs to compare with DLAs.

At a first glance there is  general consistency between  DLA, BCG,
and halo star measurements, with the exception of a  
few  cases in which the [N/Fe] are lower in DLAs.
However, the interpretation of the observed [N/Fe]
ratios must take into account that iron is one of the most refractory
elements, while nitrogen is undepleted, as we discussed in Section 4.3.
Before drawing any conclusions from the comparison of the [N/Fe]
ratios it is hence necessary to take into account the presence of dust. 

The [N/Fe] values corrected for dust depletions are shown in Fig. 7
where we use the same symbols for DLAs as in Fig. 6. 
The net effect of the correction  is a diagonal
shift of the points -- at different lengths for each object -- which 
results from the increase
in Fe abundance at invariant N abundance.

The iron depletions were estimated with two methods.
For the DLAs  which have both Fe and Zn abundance
measurements 
we estimated the amount of depletion by assuming  that
zinc tracks closely iron, [Zn/Fe] $\simeq$ 0,  as is indeed observed  in   
Galactic stars with metallicities $-2.9 \leq$ [Fe/H]
$\leq -0.2$  (Sneden, Gratton, \& Crocker, 1991) 
and in the Large Magellanic Cloud stars (Russel \& Dopita 1992), and by   
assuming  that dust in   
DLAs is of the same type as the Galactic one. 
With both assumptions we estimated the Fe and Zn depletions 
starting from the Galactic interstellar values 
$\delta$(Fe) = $-1.2$ dex and
$\delta$(Zn) = $-0.19$ dex (Savage \& Sembach, 1996, Roth \& Blades 1995)
).
This approach is similar to that employed  
by Pettini et al. (1994, 1997), with the difference that we take into
account not only iron depletion, but also that of zinc. 
The two DLAs  with both Fe and Zn measurements available
are indicated in Fig. 7 with a filled triangle
(system towards QSO 0100+1300) and a star 
(system towards QSO 1331+1700).

For the remaining DLAs without Zn but  
with Fe  measurements  available 
we estimated the correction for dust depletion from
Eq. (17) by Vladilo (1998) by adopting the average
dust-to-metals ratio in DLAs 
$< \log ( \tilde{k} / \tilde{Z} ) > = -0.21 \pm 0.16$ given there. 
The resulting [N/Fe]$_{corr}$ values are indicated in Fig. 7
with a 45 degree error bar which corresponds to the 
above quoted dispersion
in the dust-to-metals ratio. 
The left and up arrows shown in Fig. 7 for
the absorber towards QSO 0000--2620 (filled square)
were derived from 
the stringent Zn upper limit available for the system
(see Table 5), i.e. by taking [Fe/H]$_{corr}$ $\simeq$[Zn/H]$_{obs}$
and [N/Fe]$_{corr}$ $\simeq$ [N/Zn]$_{obs}$, respectively.

Since the data points are now corrected for
the presence of dust, the spread observed in Fig. 7 should 
reflect intrinsic differences
in the relative nitrogen abundance. 
Particularly relevant is the DLA at $z$ = 3.390 towards QSO
0000--2620, (filled square) for which  [N/Fe] remains
significantly higher than in the other two DLAs with zinc available.
This spread is specific to N and has no equivalent in
other abundance ratios measured in DLAs, which are  
remarkably similar both at face values  
(see e.g. Fig. 23 by Lu et al.  1996) and after 
correction for dust depletion (Vladilo 1998).
 
As can be seen in Fig. 7 most  of the [N/Fe]$_{corr}$ measurements in DLAs fall well
below those of  the metal-poor halo stars. 
The slight underabundance of the measurements taken at face value
shown  in
Fig. 6 is  in fact amplified after correction for dust depletion.  
This  conclusion still holds if we use the B star abundances, instead of the solar ones,
as a reference for  the "cosmic" abundances since,
in this case, the [N/Fe] ratio
remains unchanged and the data points in Fig. 7  shift  along 
the x-axis by only +0.18 dex. 
 
In blue compact galaxies, the [N/Fe] determinations
are based on observations of emission lines from HII regions where
dust is also known to be present  
but is not accounted for (Garnett et al. 1995).
If the presence of dust is considered, the BCGs data points of Fig. 6
would shift in the same direction as  the DLA points in Fig. 7.   
Therefore, we cannot exclude that [N/Fe] measurements
in BCGs are consistent with  DLA determinations, as
they were before the correction for the presence of dust.

After correction for dust depletion the majority of the DLA data are in line
with a pure secondary origin of nitrogen, which in Fig. 7 is sketched 
as a
dashed-line passing through the solar point.  This is particularly
clear at least in two cases: namely the DLAs towards 
QSO 0100+1300 (triangle)
and towards QSO
1331+1700 (star).  This behaviour is quite different 
from that of halo metal-poor stars
which show a constant [N/O] with metallicity,
consistent
with a primary origin of nitrogen. 
The difference between the nitrogen abundances of Galactic metal-poor stars and DLAs  
suggests that the  chemical
enrichment history of these DLAs has been somewhat different from 
that experienced by our own Galaxy.
However, there is at least one remarkable  
exception for QSO 0000-2620 where the nitrogen remains particularly
high, at the level observed in Galactic halo stars.

\subsection{The [N/$\alpha$] abundance ratio}

The [N/$\alpha$] ratio is an important  tool when unravelling the 
nucleosynthetic history 
of nitrogen.  The abundances for the $\alpha$ elements S, Si, and O
are reported in Table 5.
Unfortunately it is quite difficult
to obtain accurate
oxygen abundances in DLAs, as we have discussed in Section 3. 
This problem has often been 
circumvented by taking S or Si as a proxy for O, under the assumption
that the [$\alpha$/O] ratio does not evolve with metallicity.  
S and  Si  are $\alpha$-elements produced in the same
massive stars that produce O, and therefore the ratios between these
elements are predicted to vary little, if at all, with time (Timmes
Woosley \& Weaver 1995).  
 
However, some caution must be taken when using silicon as a proxy of oxygen.
From the theoretical point of view, type I SNae
produce more silicon than oxygen (Nomoto et al.  1984;
Thielemann et al.  1993). 
The observations indicate that Si traces O in Galactic stars 
(Kilian-Montenbruck, Gheren \& Nissen 1994; Venn 1995),
but is slightly underabundant relative to O in extragalactic HII regions
(Garnett et al. 1995). This underabundance is
interpreted as due to Si depletion onto dust grains. 
In addition to the uncertainty due to silicon depletion, we remark that
Si lines are often saturated even at the low column densities typical of DLAs. 
For these reasons  
we prefer to adopt sulphur as a tracer of the $\alpha$-elements.
Sulphur is found to track closely
oxygen in Galactic stars  (see references in
Lauroesch et al.  1996), in
dwarf galaxies  (Garnett
1989, Skillman \& Kennicutt 1993, Skillman et al.  1994) 
and in BCGs (Thuan et
al.  1995) in the entire range of metallicities explored.  
Sulphur is not depleted and, when detected, its lines are unsaturated. 
Hence, we consider it
safe to assume that [N/O] $\simeq$
[N/S] in DLAs. 
 
In Fig. 8 we compare  [N/O] measurements 
in DLA systems, in Galactic stars, and in dwarf irregular
galaxies with predictions emerged from
chemical evolution models. 
Unless stated otherwise,
DLA measurements have been derived from sulphur by
assuming  that [O/S] $\equiv$ 0. 
Larger sized symbols represent DLA measurements, and
filled symbols highlight the 
determinations obtained by our group. 
The dashed-line   represents the mean trend  of the [N/O] 
ratio in Galactic 
stars 
of all metallicities (Tomkin \& Lambert 1984).
The  dots 
represent the dwarf galaxies compiled by 
van Zee et al. (1996), van Zee L., Haynes P.M. \& Salzer J.J. (1997) and  by
Kobulnicky \& Skillman (1996)
which include the BCGs of Thuan et al.  (1995). 

The solid lines are
theoretical predictions of [N/O] from a chemical evolution model for the
solar neighbourhood under different assumptions for
the nucleosynthesis of nitrogen,  
described in detail by Matteucci, Molaro \& Vladilo (1997).  The curve
labelled "S" assumes   purely
secondary  nitrogen produced in stars of all masses. The
"P" curve assumes some amount of primary production in intermediate
mass stars  according to Renzini \& Voli (1981) 
and with a convective scale length $\alpha$=1.5.
%an efficiency of the convection parameter $\alpha$=1.5. 
The "Pmassive" curve with a plateau  assumes  primary N
both in intermediate mass stars  and in massive stars, 
where in all the other tracks it is secondary.
  
As can be seen in the figure, the [N/$\alpha$] ratios in DLAs are highly 
scattered and 
the majority of the measurements
have no precedents, at comparable level
of metallicity, in Galactic stars.
Instead, intrinsic  dispersion is  
observed in dwarf galaxies, with a range of [N/O] 
values similar
to those found in DLA galaxies.
Since both N and S 
depletions are expected to be negligible, the
new plot provides further evidence -- in addition to Fig. 7 -- that the 
dispersion of N abundances is real and
not due to different amounts of dust among the DLAs. 

The detailed study of individual cases shows peculiar behaviours
with respect to the predictions of chemical evolution models. 
The  [N/S]  ratios in the 
DLAs towards QSO 0100+1300 (triangle), QSO
1331+1700 (star), QSO 0347--3819 (octagon) and the upper limit 
towards QSO 2348-1444 (empty square with arrow)
are in between the primary and secondary nitrogen evolutionary tracks.   
The [N/S] value in the DLA towards
QSO 0930+2858 (empty square in the middle of Fig. 8)
is   close to the curves which consider a primary component of N either
in intermediate and massive stars. 
The value for the DLA towards
QSO 2343+1230 (empty square at the top of the figure) 
falls well above the
area limited by the evolutionary tracks with secondary plus
primary production of nitrogen. 
This is probably also true  for the absorber towards QSO 0000-2620. 
In this case we show in Fig. 8 both the lower limit derived from
the conservative sulphur upper limit (Lu, Sargent \& Barlow  1998;  
filled square with arrows) and 
the value from the direct oxygen measurement (Molaro et al. 1996;
filled square at the left top of the figure). 
This last determination was obtained by constraining the
$b$ value of the saturated oxygen lines from the broadening obtained from
the nitrogen lines (see discussion in Section 3).  
Even considering the uncertainty of the oxygen measurement  
it is clear that nitrogen has an intrinsically high abundance 
in this DLA.
 
It is important to note that for  
specific DLA systems  the 
[N/$\alpha$] and [N/Fe]$_{corr}$ values    
are mutually consistent. 
The same points which 
show a secondary behaviour in [N/$\alpha$] also show a secondary 
behaviour in [N/Fe]$_{corr}$. 
We stress that the  [N/Fe] and [N/S] results would not lead to
consistent results if iron  depletion were not  taken into account. 
 
The different values of N/O observed at a given O/H metallicity 
in DLAs may be attributed to a
delay between the delivery of O and N into the ISM when the star
formation proceeds in bursts (Edmunds \& Pagel 1978). 
This delay has also been  suggested by Garnett (1990) to explain the N/O scatter  
in dwarfs irregulars. According to this picture, 
galaxies which have experienced a recent episode of star formation 
show low N/O ratios due to the quick production of O in massive
stars and the prompt return to the ISM which is over after a few
10$^7$  years.  On the other hand, in galaxies that have been
quiescent for a long period, nitrogen 
would be returned to the ISM after a few 10$^8$  years 
mainly as a product of intermediate mass stars. 
During the long quiescence 
the N/O ratio will increase at about constant O/H.  
In this model, different ages or temporal gaps from recent bursts 
of star formation are responsible for the dispersion of the N/O ratios, with
low values implying a recent burst and high values pointing to a long
quiescent period.  
Skillman (1998) discussed  two cases of dwarf
irregular galaxies with  values of   [N/O] around --0.7 for which
there is clear evidence of recent star formation, in agreement with
the interpretation of a prompt O enrichment of the ISM. 
 
However, some problems arise in this scenario if DLA abundances of other
elements are considered.
If the delayed model also applies to DLAs, 
then it seems reasonable to expect  
an overabundance of the $\alpha$-elements relative to iron-peak 
elements when low values of [N/O] are observed, i.e.
in the phase temporally near the starburst. In fact in this phase  
the elemental production is dominated by Type II SNae which are
rich in $\alpha$-elements.
The few cases in which a measurement of the [$\alpha$/Fe] ratio
and of nitrogen abundances are available do not conform  with
this prediction. For instance,
as discussed in MCV, the DLA galaxies towards QSO 0100+1300 and QSO 1331+1700 
have [S/Zn] $\simeq$ 0 even if they show the lowest [N/$\alpha$] values. 
In the absorber towards QSO 1331+1700 it has been suggested
that an intervening HII region could lead to an artificially low
[N/S] ratio (see Section 4.2), since [N/O] $>$ [N/S]  
in this system (Kulkarni et al. 1996).
For the system towards QSO 0100+1300 it is improbable that the
[N/S] ratio is artificially lowered due to 
an intercepted HII region since
independent fits of the NI and SII lines 
give consistent $b$ values (see MCV), while differences between the
profiles of the two species would be expected if contribution from 
HII regions were present. 
The low [N/S] value in this DLA absorber  
is therefore difficult to understand in the framework of the N-delayed model 
since the expected overabundance of the [$\alpha$/Fe] ratio is not appreciable.

Also the   highest [N/$\alpha$]  ratios observed in the DLAs in Fig. 8 
are difficult to reconcile with the delayed nitrogen release. 
In fact in this model, N/O ratios 
are expected to lie in the region limited by the tracks of
primary and secondary nitrogen production  (Vila-Costas \& Edmunds,
1993). As we mentioned before,
the system towards QSO 2343+1230, and perhaps also the one towards
QSO 0000-2620,
lie  well above the track of primary plus secondary production instead. 
Matteucci, Molaro \& Vladilo (1997) showed that the
highest N/O values observed in DLA galaxies
can only be reproduced  by  chemical evolution models 
which assume primary production of N in massive stars, together with
selective galactic winds which carry away, preferentially, the
products of type II SNae explosions, such as oxygen 
and other $\alpha$ elements. 

We conclude that the time delay model requires other {\it ingredients} 
(for example selective galactic winds among others)  in order to
explain some of the N/O ratios observed in DLAs, once they are integrated 
with the other abundances observed in these galaxies. 
New abundance determinations are needed in order to understand how
common are the very low and extremely high ratios in DLA galaxies.

\section{Summary}

We have presented three new measurements of nitrogen abundances
in DLA galaxies, increasing by 50\% the number of determinations available
in the literature (not including upper limits). 
%In order to cast light on the chemical history  
%of DLA galaxies 
%we have investigated 
The abundances of nitrogen have been discussed  in connection with the
iron-peak elements (iron and zinc) and $\alpha$-elements
(represented by sulphur)  
for the nine absorbers with available N data. The effect of dust depletion on the 
abundance determinations
has been taken into account.   
We have compared the nitrogen measurements in DLA galaxies
with measurements   in other  astrophysical sites  
of low metallicity and with predictions from chemical evolution models. 
The main results of the present work can be summarised as follows.

\begin{itemize}

\item
The N/S and N/Fe abundance ratios show
a scatter of one order of magnitude or more,  at variance with the remarkably low
scatter observed for other elemental ratios measured in DLA galaxies.    

\item 
When the presence of dust
is considered to correct the iron abundances, 
a substantial fraction of [N/Fe] values
are lower than those in Galactic halo stars. 
A similar conclusion holds for the [N/$\alpha$] ratios obtained from 
the [N/S] measurements which are expected to be 
unaffected by dust depletion. These results suggest that the chemical evolution 
of DLA galaxies has been somewhat different from that
experienced by the Milky Way.

\item
The [N/Fe] ratios in BCGs are similar to those measured
in DLA galaxies and this similarity may remain also after correction
for the presence of dust, provided this correction is also appropriate for BCGs. 
The N/O ratios in dwarf galaxies have similar values and exhibit a similar spread
to those observed in DLA galaxies.

\item
When compared with theoretical evolution models,
it is not possible to explain the observations in DLA galaxies
by invoking a  unique  production mechanism.  Some cases are consistent with
a secondary behaviour of nitrogen, while others  
require primary production.
This may suggest that DLA absorbers include  galaxies with
different chemical histories. 
Most of the scatter of the nitrogen abundances can also be interpreted in 
the framework of the delayed model for nitrogen production
(Edmunds \& Pagel, 1978; Garnett 1990). 
However in this scenario DLA galaxies with 
low [N/$\alpha$] values, which should be temporally close to the starburst, 
are expected to show an enhancement of $\alpha$-elements which is instead
not observed, making necessary the addition of new ingredients to the model
in order to explain the observed abundance ratios.

\item
In one or possibly two DLA galaxies there is evidence for an
extremely high N/O ratio, well above the values measured
in Galactic halo stars or in extragalactic HII regions.  
The few DLA systems with
N/O ratios higher than those in any other astrophysical site 
cannot be accommodated in the framework of the delayed nitrogen model:
in this case differential galactic winds plus primary N production in
massive stars are  required
as ingredients of the chemical evolution model
(Matteucci et al. 1997). 

\end{itemize}

\clearpage

\figcaption[f1.eps]{ NI lines of the z$_{abs}$=2.3090 DLA system
in the normalised spectrum of QSO 0100+1300.
Lower panel:  1134\,\AA\ multiplet. 
Upper panel:  1200\,\AA\ multiplet.  
Smooth lines: synthetic spectrum obtained from the fit of the two bluest unblended
transitions of the 1134\,\AA\ multiplet.  
\label{fig1}}

\figcaption[f2.eps]{
NI lines of the z$_{abs}$=3.0250 DLA system 
in the normalised spectrum of QSO 0347--3819.
Lower and upper panels as in Fig. 1. 
Smooth lines: synthetic spectrum obtained by fitting the cores
of all the line profiles, with the exception of the bluest line of the
1134\,\AA\ multiplet.
\label{fig2}}

\figcaption[f3.eps]{
NI lines of the z$_{abs}$=2.8268 DLA system in
the normalised spectrum of QSO 1425+6039.
Lower and upper panels as in Fig. 1.  
Smooth lines:  synthetic spectrum obtained from the fit of the two bluest unblended
transitions of the 1134\,\AA\ multiplet.  
\label{fig3}}

\figcaption[f4.eps]{
Portions of the spectrum of QSO 0347--3819
including FeII, OI, SII and SiII lines of the z$_{abs}$=3.025 DLA system.
Rest wavelengths of the transitions are indicated in each panel.  
Radial velocities are relative to the redshift of the main component
z$_{abs}$=3.0247.   
Smooth lines: synthetic spectra derived as explained in Section 3.2. 
\label{fig4}}

\figcaption[f5.eps]{
Portions of the spectrum of QSO 1425+6039 including FeII, OI, SII and SiII lines of the 
DLA system at z$_{abs}$=2.8268. 
Radial velocities are relative to this redshift value. 
Smooth lines:  synthetic spectra obtained as explained in Section 3.3. 
\label{fig5}}

\figcaption[f6.eps]{
Measurements of the [N/Fe] ratio in DLA systems.
Large filled symbols: determinations obtained by our group;
square: DLA system towards QSO 0000--2619;
triangle: QSO 0100+1300; octagon: QSO 0347$-$3819; diamond: QSO 1425+6039.
Large empty symbols: data taken from literature;
star:  DLA system towards QSO 1331+1700. 
Region between dotted lines: range of [N/Fe] values 
measured in Galactic metal-poor stars by Carbon et al. (1987); 
solid line: linear regression to these stellar data. 
Small circles: [N/Fe] values measured in blue compact galaxies by Thuan et al. (1995).
Dashed line: expected behaviour of the [N/Fe] ratio with metallicity for
secondary nucleosynthetic origin of nitrogen.  Sun symbol: solar reference abundance.  
\label{fig6}}

\figcaption[f7.eps]{
Measurements of the [N/Fe] ratio in DLA systems corrected for dust depletion.
DLA systems are indicated with the same symbols as in Fig. 6.
In two cases (triangle and star) the dust depletion is estimated from
available Fe and Zn abundance determinations. In the other cases, indicated with
diagonal error bars, the dust depletion is estimated from Fe
measurements and from the typical range of dust-to-metals ratios in DLA systems.
See Section 5.1 for more details.  
Dotted lines, solid line, and dashed line as in Fig. 6. 
\label{fig7}}

\figcaption[f8.eps]{
Measurements of the [N/$\alpha$] ratio in DLA systems and
other astrophysical sites of low metallicity.
DLA systems are indicated with the same symbols as in Fig. 6.
The [N/O] values in DLA systems are obtained from sulphur abundance determinations
by assuming [S/O]$\equiv$0. The filled square top left corresponds to the
direct O determination for the system towards QSO 0000--2619 by Molaro et al. (1996). 
Dashed-line: mean trend of [N/O] values observed in Galactic halo and disk stars 
by Tomkin \& Lambert (1984).  
Dots:  [N/O] measurements in HII regions in dwarfs irregular galaxies
compiled by Kobulnicky \& Skillman (1996) and van Zee et al. (1996, 1997). 
Solid lines: evolutionary tracks of [N/O] versus [O/H] computed by  
Matteucci et al. (1997). See Section 5.2 for more details.
\label{fig8}}

\clearpage

\begin{deluxetable}{lcccc}
%\doublespace
\footnotesize
\tablecaption{Journal of observations} 
\tablewidth{0pt}
\tablehead{
\colhead{QSO}     & 
\colhead{Date}   &
\colhead{t$_{exp}$ (s)}&
\colhead {No. spectra} &
\colhead {Coverage (\AA)}}
\startdata
0100+1300& 22-9-1996  & 10800 & 3 & \nl
         &            &       &   & 3657-4617 \nl
         & 03-12-1996 & 10800 & 3 & \nl
           
0347$-$3819 & 17-09-1995 &  12000 & 2  & 4310-5600 \nl

1425+6039 & 23-7-1996 & 5400 & 3 & \nl
          &           &      &   & 4107-5617 \nl
          & 05-8-1996 & 4800 & 2 & \nl
\enddata
\end{deluxetable}

\clearpage
\begin{deluxetable}{lcl}
%\footnotesize 
\tablecaption{Transitions under study\tablenotemark{a}}
\tablewidth{0pt}
\tablehead{
\colhead{Ion} &
\colhead{Vacuum rest} &
\colhead{ f$_{\lambda}$  }  \nl
\colhead{} &
\colhead{wavelength (\AA)} & \colhead{} \nl
}
\startdata
NI  & 1134.1653    & 0.01342  \nl
    & 1134.4149 & 0.02683    \nl   
    & 1134.9803 & 0.04023    \nl

NI  & 1199.5496  &  0.13280     \nl
    & 1200.2233 & 0.08849     \nl
    & 1200.7098 & 0.04423      \nl

FeII &1081.8748 & 0.01400   \nl
 & 1096.8769  & 0.03199  \nl
 & 1125.4477  & 0.01099     \nl
 & 1127.0984  & 0.00300       \nl
 & 1133.6650  & 0.00600   \nl
 & 1143.2260  & 0.01331       \nl
 & 1144.9379  & 0.10500      \nl
 & 1260.5330  & 0.02500      \nl

SiII & 1190.4158   & 0.25020       \nl
 & 1193.2897  & 0.49910        \nl
 & 1260.4221  & 1.00700        \nl
 & 1304.3702  & 0.09400        \nl
 
SII & 1250.5840   & 0.00545     \nl
 & 1253.8110  & 0.01088       \nl
 & 1259.5190  & 0.01624       \nl

OI &  1302.1685   & 0.04887       \nl
 &    1355.5977  & 1.248E-06      \nl
\enddata        

\tablenotetext{a}{Laboratory vacuum wavelengths and
oscillator strengths are taken from Morton (1991)
with the only exception of the SiII 1304\AA\ transition
for which the values by Tripp et al. (1996) are adopted. }
   
\end{deluxetable}

\clearpage

\begin{deluxetable}{lcccccc}
\footnotesize
\tablecaption{Nitrogen determinations in DLA systems}
\tablewidth{0pt}
\tablehead{
\colhead{QSO}        & \colhead{$z_{\rm abs}$} & 
\colhead{log N(HI)}  & \colhead{ Ref.}         & 
\colhead{log N(NI) } & \colhead{b}             & \colhead{ Ref.} 
}
\startdata
0000$-$2620 & 3.3901& 21.40$\pm$0.10   & 1 & 14.68$\pm$0.14  &13.7$\pm$1.8& 2 \nl
            &       & 21.41$\pm$0.08   & 9\nl

0100+1300 & 2.3090  & 21.40$\pm$0.05    & 4 & 14.69$\pm$0.07  &11.2$\pm$1.0 & 3 \nl
          &         & 21.32\tablenotemark{a}& 12 &15.29$\pm$0.35\tablenotemark{b}\nl
          
0347$-$3819 & 3.0250& 20.70$\pm$0.10    & 5 & 14.60$\pm$0.05  &17.9$\pm$1.1 & 3 \nl

0930+2858 &3.2353&20.18\tablenotemark{a}& 12& 13.82\tablenotemark{b}  & ...          &
12\nl

1055+4611 &3.3172&20.34\tablenotemark{a}&12 & $\leq$ 14.09\tablenotemark{b}    & ...&
12\nl

1202-0725 &4.3829&20.60$\pm$0.12        &13 & $\leq$ 14.30\tablenotemark{b}    & ...&
12\nl

1331+1700 & 1.7764  & 21.18$\pm$0.05    & 6 & 14.5$\pm$0.10 & \tablenotemark{m}&7\nl
          &         &                   &   & 14.53\tablenotemark{a}&
          \tablenotemark{m}&8 \nl

1425+6039 & 2.8268  & 20.30$\pm$0.04    & 9 & 14.70$\pm$0.04  & 23.00$\pm$1.8& 3 \nl
 
1946+7658 & 2.8443  & 20.27$\pm$0.06    & 9 & 12.53\tablenotemark{b}     & ...        
 & 12  \nl

2212-1626 & 3.6617  & 20.20$\pm$0.08    & 9 &$\leq$ 13.58\tablenotemark{b}    & ...   
 & 12\nl

2233+1310 & 3.1493&20.00\tablenotemark{a}&12&$\leq$ 14.32\tablenotemark{b}    & ...   
 & 12\nl

2237-0608 & 4.0803&20.52$\pm$0.11       &9  &$\leq$ 14.28\tablenotemark{b}    & ...   
 & 12\nl

2343+1230 & 2.4313&20.34\tablenotemark{a}&12& 14.67\tablenotemark{b} & ...            
 & 12  \nl

2344+124  & 2.5379  & 20.43$\pm$0.09    & 10 & $<$ 13.81       & ...          & 10 \nl
          &       &20.36\tablenotemark{a}&12& 13.55\tablenotemark{b}  &              &
          12 \nl
          
2348--1444& 2.2794  & 20.57$\pm$0.09    & 5 & $\leq$ 13.47       & ...          &
11\nl
\enddata
\tablenotetext{a}{Error bar not reported by the author}
\tablenotetext{b}{N(NI) obtained from [N/H] value reported by the authors 
who used log\, (N/H)$_{\sun}$ = --3.95$\pm$0.04}
\tablenotetext{m}{Multiple components}
\tablenotetext{}{ 
REFERENCES: (1) Savaglio et al. (1994); (2)  Molaro et al. (1996); (3) This paper;
(4) Pettini, Boksenberg \& Hunstead (1990); (5) Pettini et al. (1994); 
(6) Wolfe (1995b); (7) Green et al. (1995);
(8) Kulkarni et al (1996) ; (9) Lu et al. (1996); 
(10) Lipman 1995; (11) Pettini, Lipman \& Hunstead (1995);
(12) Lu, Sargent \& Barlow (1998); (13) Lu et al. (1996b)
}
\end{deluxetable}

\clearpage
\small
\begin{deluxetable}{lccccccccc}
%\doublespace
\footnotesize
\tablewidth{-2cm}  
\tablecaption{Column densities of other species.} 
%\tablewidth{0pt}
\tablehead{
\colhead{QSO}    & \colhead{$z_{\rm abs}$} & 
\colhead{log N(SII)}  & b&
\colhead{log N(SiII)} &b&
\colhead{log N(OI)}  &b&
\colhead{log N(FeII)} &b
} 
\startdata
 0347$-$3819 & 3.02476& 14.77$\pm$0.08& 17.9$\pm$3.3 & 15.07$\pm$0.43
& 21$\pm$5   &16.07$^{+0.90}_{-0.46}$& 17.9$\pm$3.3  & 14.35 $\pm$0.10& 17.9 \nl
             &       &               &              &
&            & $\leq$ 17.95\tablenotemark{a}  &               &                &       \nl

1425+6039   & 2.82680 & ...*       & ...            & $\geq$14.32 & 23.0 &
 ...*        & ...  & 14.45$\pm$0.06& 25.2$\pm$1.8 \nl
            &      &               &               &            &      &
$\leq$17.85\tablenotemark{a} &      &               &               \nl 
\enddata
\tablenotetext{a}{From the upper limit on the OI 1355 line}
\tablenotetext{*}{Not detected due to contamination by the Ly$\alpha$ forest}
\end{deluxetable}

\clearpage
%\ptlandscape 
\begin{deluxetable}{lcccccccc}
\doublespace
\footnotesize
\tablewidth{-1.5cm}  
\tablecaption{Elemental abundances in DLA systems with nitrogen measurements\tablenotemark{a,b,c}.} 
%\tablewidth{0pt}
\tablehead{
\colhead{QSO}    & \colhead{$z_{\rm abs}$} & 
\colhead{[N/H]}  & 
\colhead{[S/H]}  &
\colhead{[Si/H]} &
\colhead{[O/H]}  &
\colhead{[Fe/H]} &
\colhead{[Zn/H]} &  
\colhead{ Refs.}       
} 
\startdata
0000$-$2620 & 3.3901& --2.69$\pm$0.19& $ \leq$--1.97 &  --2.48$\pm$0.19 &
--3.07$\pm$0.18 
& --2.41$\pm$0.16  & $\leq$ -1.90       &1,2,3 \nl

0100+1300 & 2.3090  & --2.68$\pm$0.11& --1.60$\pm$0.09&  --1.36$\pm$0.05 & $\geq$
--2.83 
& --2.02$\pm$0.11  & --1.55$\pm$0.11    & 3,4,5,15 \nl

0347$-$3819 & 3.0250& --2.07$\pm$0.13& --1.20$\pm$0.14  & --1.18$\pm$0.44 &
--1.50$^{+0.91}_{-0.47}$ 
& --1.86$\pm$0.14  & $\leq$--0.83               & 5,14 \nl

0930+2858   &3.2353& --2.33$\pm$0.15& --1.78$\pm$0.15 & $\geq$ --1.79    &$\geq$
--2.51
&  ...             &  ...                & 3 \nl

1055+4611   &3.3172& $\leq$ --2.22  & $\leq$ --1.26    & $\geq$ --1.60    &$\geq$
--2.09
&  ...              &    ...               &  3 \nl

1202$-$0725 &4.3829&  $\leq$ --2.27  & ...          & --1.76$\pm$0.12  & $\geq$ --1.96

&  -2.23$\pm$0.16    &  ...             &  3,13 \nl

1331+1700 & 1.7764 & --2.65$\pm$0.13  & --1.35$\pm$0.12 & --1.85$\pm$0.10 & 
--2.74$\pm$0.31
& --2.05$\pm$0.10   & --1.27$\pm$0.06  &7,8,9\nl

1425+6039   & 2.8268& --1.57$\pm$0.09& ...             & $\geq$--0.95      &   $<$
0.68    
& --1.36$\pm$0.07  & ...                & 5,6\nl

1946+7658 & 2.8443&  -3.71$\pm$0.10  & $\leq$ --1.79    & --2.09$\pm$0.06& $\geq$
--2.69
& -2.40$\pm$0.07  & $\leq$--0.82         & 3,10  \nl

2212$-$1626& 3.6617& $\leq$--2.59     & ...             & --1.90$\pm$0.08  &
$\geq$--2.30     
&  $\leq$--1.78             &  ...                  &  3,6 \nl

2233+1310  & 3.1493 & $\leq$--1.65     & ...             & $\geq$--1.04     &
$\geq$--1.49
&  ...                 &  ...                 &  3 \nl

2237$-$0608& 4.0803 & $\leq$--2.21     & ...             & --1.80$\pm$0.11  & ...
&  -2.18$\pm$0.16      &   ...                &  3,6 \nl

2343+1230  & 2.4313 & --1.64$\pm$0.10  & --1.90$\pm$0.08 & --0.74$\pm$0.11  &
$\geq$--1.86
&   ...                &    ...               &  3 \nl

2344+124  & 2.5379 & --2.78$\pm$0.14    & $\leq$ --1.30   & $\geq$--1.67      &
$\geq$--2.26
& --1.78\tablenotemark{d} (11)          & ...  &  3,11  \nl

2348--1444& 2.2794 & $\leq$--3.07       & --1.91$\pm$0.19   & --1.97$\pm$0.15     &
$\geq$--2.64
&  --2.35$^{+0.24}_{-0.13}$       & $<$--1.31   & 12 \nl
\enddata
\tablenotetext{a}{All abundances have been normalised to the meteoritic
values reported by Anders \& Grevesse (1989), with the
exception of N and O for which the most recent solar phostospheric values   
given by Grevesse \& Noels (1993) have been adopted: 
log (N/H)$_{\sun}$=--4.03$\pm$0.07, log (S/H)$_{\sun}$=--4.73$\pm$0.05,
log (Si/H)$_{\sun}$=--4.45$\pm$0.02,log (O/H)$_{\sun}$=--3.13$\pm$0.07,
log (Fe/H)$_{\sun}$=--4.49$\pm$0.01, log (Zn/H)$_{\sun}$=--7.35$\pm$0.02}
\tablenotetext{b}{Errors in [X/H]=log (X/H)$_{obs}$--log (X/H)$_{\sun}$ 
include errors 
in X and H column densities and errors in solar abundances.}  
 
\tablenotetext{c}{The literature data have been corrected to the solar values
given in (a)}
\tablenotetext{d}{Error bar not reported by the author}

\bigskip
 
\tablenotetext{}{ 
REFERENCES: (1) Molaro et al. 1996; (2) Pettini et al. (1997b); (3) Lu, Sargent \&
Barlow 1998; 
(4) Molaro, Centuri\'on \& Vladilo (1997); (5) This paper; (6) Lu et al. (1996);
(7) Green et al. (1995); (8) Kulkarni et al. (1996); (9) 
Wolfe (1995b);
(10) Lu et al. (1995); (11) Lipman (1995); (12) Pettini, Lipman \& Hunstead (1995);
(13) Lu et al. 1996b;
(14) Pettini et al. 1994; (15) Wolfe et al. (1994)
} 
\end{deluxetable}

\end{document}